\documentclass[prd,twocolumn,superscriptaddress,preprintnumbers,balancelastpage,nofootinbib,floatfix]{revtex4-2}

\usepackage{graphicx}  % needed for figures
\usepackage{dcolumn}   % needed for some tables
\usepackage{bm}        % for math
\usepackage{amssymb, amsmath}
\usepackage{slashed}   % for math
\usepackage{lipsum}
\usepackage[usenames,dvipsnames,svgnames,table]{xcolor}
\usepackage{comment}
\usepackage[utf8]{inputenc}
\usepackage{multirow}
\usepackage{subfigure}
\usepackage{xfrac} % Nice small fractions \sfrac

\definecolor{BlueViolet}{rgb}{0.2, 0.00, 0.7}
\definecolor{Blue}{rgb}{0.15, 0.00, 0.9}
\definecolor{light-blue}{rgb}{0.3, 0.3, 1}
\definecolor{kit-green}{rgb}{0, 
0.58823 %150/255
, 0.50980 %130/255
}
\usepackage[%dvipdfmx,
colorlinks=true, linkcolor=light-blue,citecolor=light-blue,urlcolor=kit-green]{hyperref} 

\newcommand{\bea}{\begin{eqnarray}}
\newcommand{\eea}{\end{eqnarray}}

\def\Kbar    {\kern 0.18em\overline{\kern -0.18em K}{}\xspace}

\usepackage{adjustbox}
\usepackage{tabularx}
\newcolumntype{Y}{>{\centering\arraybackslash}X} %for tabularx
\usepackage{booktabs} %for \toprule
\usepackage{colortbl} %for \rowcolors
\def\beq#1\eeq{\begin{align}#1\end{align}}

\RequirePackage{xspace}
\def\Bbar    {\kern 0.18em\overline{\kern -0.18em B}{}\xspace}

\begin{document}
%%%%%%%%%%%%%%%%%%%%%%%%%%%%%%%%%
\preprint{CHIBA-EP-278}

\title{Exact baryon-meson relation for \texorpdfstring{\boldmath{$b \to s \nu\bar\nu$}}{b to s nu nubar}}

\author{Teppei Kitahara}
\email{kitahara@chiba-u.jp}
\affiliation{
Department of Physics, Graduate School of Science,
Chiba University, Chiba 263-8522, Japan}
\affiliation{
  Kobayashi-Maskawa Institute for the Origin of Particles and the
  Universe, Nagoya University,
  Furo-cho Chikusa-ku, Nagoya 464-8602, Japan
}

\author{Manas Kumar Mohapatra}
\email{manasmohapatra12@gmail.com}
\affiliation{School of Physics, University of Hyderabad, Hyderabad, India - 500046}

\author{Kota Sasaki}
\email{sasaki@chiba-u.jp}
\affiliation{
Department of Physics, Graduate School of Science,
Chiba University, Chiba 263-8522, Japan}

%%%%%%%%%%%%%%%%%%%%%%%%%%%%%%%%%%
%%%%%%%%%%%%%%%%%%%%%%%%%%%%%%%%%%
\begin{abstract} 
We derive an exact relation
among the 
normalized branching fractions of $\Lambda_b \to \Lambda \nu \bar\nu$ and $B \to K^{(\ast)} \nu\bar\nu$, 
within the lepton-number-conserving effective Hamiltonian in which only active neutrinos contribute to the invisible final state.
Although the effective Hamiltonian contains 18 independent Wilson coefficients, the three normalized rates depend on only two independent combinations of them, leading to an exact algebraic relation. 
Remarkably, 
the central values of the coefficients of this baryon-meson relation are close to $1/4$ and $3/4$, which appear in the heavy-to-heavy $b\to c$ semileptonic sum rule among the branching fractions of $\Lambda_b \to \Lambda_c \tau \bar\nu$ and $B \to D^{(\ast)}\tau\bar{\nu}$.
Once the decay rate of $B \to K^{\ast} \nu \bar\nu$ is measured, the decay rate of  $\Lambda_b \to \Lambda \nu\bar\nu$ 
can be determined in a model-independent manner for new-physics scenarios involving only 
active neutrino interactions, thereby providing a clean target prediction for future experiments.
This clearly demonstrates that observables in baryonic and mesonic $b \to s \nu \bar{\nu}$ transitions will serve as a 
compact consistency test 
and a powerful probe for discriminating among new-physics scenarios.
\\
%--------------------------------------------------\\
%{\sc Keywords: Bottom Quark, Semi-Leptonic Decays}
\end{abstract}

%%%%%%%%%%%%%%%%%%%%%%%%%%%%%%%%%
\maketitle

%%%%%%%%%%%%%%%%%
%%%%%%%%%%%%%%%%%%%%%%%%%%%%%%%%%%%%%

%%%%%%%%%%%%%%%%%%%%%%%%%%%%%%%%%%%%%
\section{Introduction}
\label{sec:intro}
%%%%%%%%%%%%%%%%%%%%%%%%%%%%%%%%%%%%%

Due to the Glashow-Iliopoulos-Maiani mechanism \cite{Glashow:1970gm}, 
rare decays induced by the $b \to s$ flavor-changing neutral current (FCNC) 
provide sensitive probes of short-distance new physics beyond the Standard Model (SM). 
The consistent tensions observed in $b \to s \ell^+ \ell^-\,(\ell=e,\mu)$ transitions \cite{Alguero:2022wkd,Alguero:2023jeh,Capdevila:2023yhq,Hurth:2023jwr}
further motivate complementary probes in channels with neutrinos in the final state \cite{Buras:2014fpa}. 
Among them, the $b \to s \nu \bar{\nu}$ modes are theoretically especially clean because, apart from the form-factor uncertainty, they are free from long-distance hadronic contributions \cite{Altmannshofer:2009ma,Bartsch:2009qp}. 
In particular, $B \to K^{(*)}\nu\bar{\nu}$
provides an important benchmark for testing new physics.

Belle II Collaboration has reported the first measurement of $B^{+} \to K^{+} \nu\bar\nu$, obtaining
$\mathcal{B}(B^{+} \to K^{+} 
\nu\bar\nu) 
=(2.3 \pm 0.7)\times10^{-5}$,
which corresponds to a $2.7\sigma$ excess over the SM prediction \cite{Belle-II:2023esi}.
Combining with the Belle \cite{Belle:2013tnz,Belle:2017oht} and BaBar results \cite{BaBar:2010oqg,BaBar:2013npw}, 
the world average is $\mathcal{B}(B^{+} \to K^{+}  \nu\bar\nu) 
=(1.3 \pm 0.4)\times10^{-5}$ \cite{Belle-II:2023esi}, which leads to
\beq
\frac{\mathcal{B}\left(B^{+} \to K^{+}  \nu\bar\nu\right) }{\mathcal{B}\left(B^{+} \to K^{+}  \nu\bar\nu\right)_{\rm SM} } =
2.82 \pm 0.88\,. 
\label{eq:BelleIIresult}
\eeq
Thus, the current data can still be regarded as consistent with the SM prediction at the $2\sigma$ level.
In contrast, searches for $B \to K^{*}\nu\bar{\nu}$ by Belle Collaboration yield 90\% CL upper bounds only \cite{Belle:2017oht}:
\beq
\mathcal{B}\left(B^0 \to K^{*0}\nu\bar{\nu}\right) &< 1.8 \times 10^{-5}\,,
\label{eq:Kstbound}\\
\mathcal{B}\left(B^+ \to K^{*+}\nu\bar{\nu}\right) &< 6.1  \times 10^{-5}\,,
\eeq
 while the corresponding Belle II result is still pending.

Beyond the mesonic sector, 
the baryonic decay $\Lambda_b \to \Lambda \nu \bar{\nu}$ offers complementary information of the $b \to s \nu \bar\nu$ FCNC because of its distinct hadronic structure. 
Several theoretical studies have explored this mode within low-energy effective Hamiltonian frameworks \cite{Chen:2000mr,Altmannshofer:2025eor}, including scenarios with left- and right-handed neutrinos~\cite{Das:2025zrn}.
It has also been analyzed within
the Standard Model Effective Field Theory~\cite{Das:2023kch}, 
as well as  in scenarios with heavy mediators~\cite{Lee:2025kvf}, invisible light particles~\cite{Hu:2025zua}, 
and other new physics contributions~\cite{Aliev:2007rm, Sirvanli:2007yq}.
These analyzes show that baryonic channels can provide  powerful and complementary constraints on new-physics scales and operator structures. In addition, rare $c \to u \nu \bar{\nu}$ and $b \to q \nu \bar{\nu}$ ($q = d, s$) FCNC transitions in both meson and baryon sectors have been investigated as sensitive null tests of the SM, with comprehensive model-independent analyses and global fits presented in Refs.~\cite{Bause:2020xzj,Bause:2021cna,Gisbert:2022lvn}.

Although no experimental measurement of $\Lambda_b \to \Lambda \nu \bar{\nu}$ is currently available, 
it is well motivated to ask whether baryonic and mesonic $b \to s \nu\bar{\nu}$ observables can be related in a robust and model-independent way.
This possibility is also motivated by the baryon-meson sum rule established for $b \to c$ semileptonic decays \cite{Blanke:2018yud,Blanke:2019qrx,Fedele:2022iib,Duan:2024ayo,Endo:2025fke,Endo:2025cvu,Endo:2025lvy,Endo:2025set,Iguro:2026xgi}.
In this work,
we derive 
an exact baryon-meson relation among the normalized branching fractions of $\Lambda_b \to \Lambda \nu \bar\nu$ and $B \to K^{(*)}\nu\bar\nu$, 
assuming that the invisible final states are active neutrinos described the lepton-number-conserving effective Hamiltonian. 
Therefore, once the decay rate of $B \to K^{*}\nu\bar\nu$ is measured, 
the decay rate of $\Lambda_b \to \Lambda \nu\bar\nu$ can be determined in a model-independent manner within this effective Hamiltonian.
This demonstrates that baryonic and mesonic $b \to s \nu \bar{\nu}$ observables provide a 
powerful consistency test 
and thereby help discriminate among new-physics scenarios.

Experimentally, $B \to K^{(*)}\nu\bar{\nu}$ modes are being actively explored, and future measurements, especially of $B^0 \to K^{*0}\nu\bar{\nu}$ at Belle II \cite{Jia:2023upb}, will provide essential input for such tests. 
By contrast, the baryonic mode $\Lambda_b \to \Lambda \nu \bar{\nu}$ remains much more challenging to access directly.
Belle II cannot produce $\Lambda_b$ baryons, while
hadron-collider experiments such as LHCb 
can produce them copiously and enable precision measurements in several channels \cite{LHCb:2023wbo}.
Nonetheless, 
their poor missing-energy resolution renders measurements of invisible final states challenging.
This makes future high-luminosity $e^+e^-$ colliders at the $Z$ pole, 
such as FCC-ee and CEPC, 
particularly attractive for this mode \cite{CEPCStudyGroup:2018ghi, FCC:2018evy, Bernardi:2022hny, CEPCPhysicsStudyGroup:2022uwl, Amhis:2023mpj}. 
In the same environment, $B_s \to \phi \nu \bar{\nu}$ is also expected to be probed with comparable or even better sensitivity to the $b \to s \nu\bar\nu$ transitions \cite{Rajeev:2021ntt,Amhis:2023mpj}.

%%%%%%%%%%%%%%%%%%%%%%%%%%%%%%%%%%%%%
\section{Theoretical Framework}
\label{sec:framework}
%%%%%%%%%%%%%%%%%%%%%%%%%%%%%%%%%%%%%
In order to study possible new-physics effects in the $b\to s \nu_i \bar{\nu}_j$ transition,
we adopt the following

low-energy effective Hamiltonian:
\begin{equation}
\begin{aligned}
\mathcal{H}_{\rm eff}
&= -\frac{4 G_F}{\sqrt{2}} V_{ts}^{\ast} V_{tb}
\frac{\alpha (M_Z)}{4\pi} C_L^{\rm SM}
\sum_{i,j=1}^{3} 
\Big[
(\delta^{ij}+C_L^{ij})\,\mathcal{O}_L^{ij}  \\
&\qquad + C_R^{ij}\,\mathcal{O}_R^{ij}
\Big]
+ \text{h.c.}\,,
\label{eq:Heff}
\end{aligned}
\end{equation}
with
\begin{equation}
\begin{aligned}
\mathcal{O}_L^{ij} &= \left(\overline{s} \gamma_\mu P_L b\right)
\left(\bar{\nu}_i \gamma^\mu (1-\gamma_5) \nu_j\right)\,, \\
\mathcal{O}_R^{ij} &= \left(\overline{s} \gamma_\mu P_R b\right)
\left(\bar{\nu}_i \gamma^\mu (1-\gamma_5) \nu_j\right)\,.
\label{eq:O}
\end{aligned}
\end{equation}
$P_{L,R} = (1\mp\gamma_5)/2$ represent the left- and right-handed projection operators, respectively.
The indices $i,j=1,2,3$ are the neutrino generations and 
the matrix $V$ denotes the Cabibbo--Kobayashi--Maskawa (CKM) matrix~\cite{Cabibbo:1963yz, Kobayashi:1973fv}. 
For the numerical value of the CKM components,
we used $\left|V_{t b} V_{t s}^*\right|=0.0406 \pm 0.0009$, which are fitted  without using the new physics sensitive observables
\cite{Allwicher:2024ncl,Crivellin:2025qsq,UTfit:2022hsi}.
The SM Wilson coefficient (WC) is given by \cite{Fael:2025xmi}
\begin{equation}
C_L^{\rm SM} = - \frac{X_t}{\sin^2 \theta_W(M_Z)}\,, \quad 
X_t = 1.471 \pm 0.012 \,, 
\end{equation}
which includes the next-to-leading order (NLO) QCD \cite{Misiak:1999yg, Buchalla:1998ba} and  NLO electroweak corrections \cite{Brod:2010hi}. 
Following Refs.~\cite{Brod:2010hi,Fael:2025xmi}, we use the six-active flavor value of the weak-mixing angle, $\sin^2 \theta_W (M_Z) = 0.23140(6)$ \cite{ParticleDataGroup:2026mpi}.
For the fine-structure constant, 
we use the latest value, $\alpha(M_Z)^{-1} \simeq 127.96(1)$ \cite{ParticleDataGroup:2026mpi}.
The effects of new physics are parameterized through 18 independent WCs $C_{L,R}^{ij}$ (all of that vanish in the SM limit). 
Since all operators are conserved in both QCD and QED, the WCs do not have scale dependence. 
Throughout this analysis, we assume that all neutrinos appearing in $\mathcal{H}_{\rm{eff}}$ are left-handed.
Note that in contrast to a convention adopted in the previous literature, 
we define $\mathcal{H}_{\rm eff}$
with $C_L^{\rm SM}$ factorized out in Eq.~\eqref{eq:Heff}, following the analysis of $b \to c$ semileptonic decays.
From a mathematical point of view, however, this is just convention of new physics sector, and the final results of our analysis remain unchanged.
Our relation is derived within effective Hamiltonian in Eq.~\eqref{eq:Heff}. 
It could be violated if additional invisible particles, right-handed neutrinos, or lepton-number-violating operators contribute.

By using this effective Hamiltonian, 
we obtain the following $b \to s \nu \bar\nu$ 
signal strengths with respect to the SM prediction for the branching fractions\footnote{In several references~\cite{Buras:2014fpa,Calibbi:2015kma,Saad:2020ucl,Bednyakov:2023njo,Buras:2024mnq}, this ratio is denoted by $R_{\nu\nu}^{K^{(\ast)}}$.} and a longitudinal polarization fraction $F_L(K^\ast)$:
\begin{widetext}
\begin{alignat}{3}
\mu(B \to K \nu \bar\nu)& =&\,\frac{\mathcal{B}\left(B^{0,+} \to K^{0,+} \nu \bar\nu\right)}{\mathcal{B}\left(B^{0,+} \to K^{0,+} \nu \bar\nu\right)_{\rm SM}} & = \frac{1}{3}\Biggl(
\sum_{i=1}^{3}  \left| 1 + C_L^{ii} + C_R^{ii}\right|^2  + \sum_{i\neq j}^{3} \left| C_L^{ij} + C_R^{ij}\right|^2\Biggr)\,,
\label{eq:BtoK}
\\
\mu(B \to K^\ast \nu \bar\nu) & =&\, \frac{\mathcal{B}\left(B^0 \to K^{\ast 0}  \nu \bar\nu\right)}{\mathcal{B}\left(B^{\ast 0} \to K^0 \nu \bar\nu\right)_{\rm SM}} ~& = \frac{1}{3}\Biggl[
\sum_{i=1}^{3}  \left( \left| 1 + C_L^{ii} \right|^2 + \left|C_R^{ii}\right|^2  - 1.18(10) \text{Re}\!\left[\left(1 + C_L^{ii}\right) C_R^{\ast ii}\right]\right) 
\nonumber \\ 
&&& \quad + \sum_{i\neq j}^{3} \left( 0.20(2) \left| C_L^{ij} + C_R^{ij}\right|^2 +0.80(2)   \left| C_L^{ij} - C_R^{ij}\right|^2\right)\Biggr]\,,
\label{eq:BtoKst}
\\
\mu(\Lambda_b \to \Lambda \nu \bar\nu) & =&\frac{\mathcal{B}\left(\Lambda_b \to \Lambda  \nu \bar\nu\right)}{\mathcal{B}\left(\Lambda_b \to \Lambda  \nu \bar\nu\right)_{\rm SM}} ~ & = \frac{1}{3}\Biggl[
\sum_{i=1}^{3}  \left( \left| 1 + C_L^{ii} \right|^2 + \left|C_R^{ii}\right|^2  - 0.37(14) \text{Re}\!\left[\left(1 + C_L^{ii}\right) C_R^{\ast ii}\right]\right) \nonumber 
\\
&&& \quad + \sum_{i\neq j}^{3} \left( 0.41(4) \left| C_L^{ij} + C_R^{ij}\right|^2 +0.59(4)   \left| C_L^{ij} - C_R^{ij}\right|^2\right)\Biggr]\,,\\
\mu(F_L(K^\ast))& = & \frac{F_L(B^0 \to K^{\ast 0} \nu\bar\nu)}{F_L(B^0 \to K^{\ast 0} \nu\bar\nu)_{\rm SM}} &= \frac{1}{3}\frac{1}{ \mu(B \to K^\ast \nu \bar{\nu})} \Biggl(
\sum_{i=1}^{3}  \left| 1 + C_L^{ii} - C_R^{ii}\right|^2  + \sum_{i\neq j}^{3} \left| C_L^{ij} - C_R^{ij}\right|^2\Biggr)\,,
\label{eq:Lambdab}
\end{alignat}
\end{widetext}
where the uncertainties comes from the form factors. 
For the form factors, 
we adopt the results obtained from refined fits
for  
$B\to K$ \cite{Gubernari:2023puw}, $B \to K^\ast$ \cite{Gao:2024vql}, and $\Lambda_b \to \Lambda$ transitions \cite{Detmold:2016pkz}, whose form factors are parametrized by 
Bharucha-Straub-Zwicky (BSZ)
$z$-series expansion with $N=3$ \cite{Bharucha:2015bzk}.
Details of the uncertainty calculation are given in the Appendix \ref{sec:app}.

 The updated SM predictions are
\beq
\mathcal{B}\left(B^0 \to K^0 \nu \bar\nu\right)_{\rm SM} &= 
\left(4.25\pm 0.15\pm 0.20\right)\times 10^{-6}\,,\nonumber \\
\mathcal{B}\left(B^0 \to K^{\ast 0} \nu \bar\nu\right)_{\rm SM} &= \left(7.87\pm0.88\pm0.37\right)\times 10^{-6}\,,\nonumber \\
\mathcal{B}\left(\Lambda_b \to \Lambda \nu \bar\nu\right)_{\rm SM} &= \left(7.97\pm 0.93\pm0.38\right)\times 10^{-6}\,,\nonumber \\
F_L\left(B^0 \to K^{\ast 0} \nu\bar\nu\right)_{\rm SM} & = 0.44 \pm 0.04\,,
\label{eq:SMexpectation}
\eeq
where the first uncertainty arises from the form-factor determinations, 
while the second reflects the other uncertainties (dominated by CKM).\footnote{For the charged modes, we obtain
$ \mathcal{B}\left(B^+ \to K^+ \nu \bar\nu\right)_{\rm SM} = 
\left(4.62\pm0.16\pm0.22\right)\times 10^{-6} 
$ and $
\mathcal{B}\left(B^+ \to K^{\ast +} \nu \bar\nu\right)_{\rm SM} = \left(8.61\pm0.95\pm0.41\right)\times 10^{-6},
$ 
where triply Cabibbo-suppressed tree-level contributions, $B^+ \to \tau^+ (\to K^{(\ast)+} \bar{\nu}_\tau) \nu_\tau$,  are not included \cite{Becirevic:2023aov}. 
In the Belle II analysis \cite{Belle-II:2023esi}, the tree-level processes are included in the background model, not in the signal.
We also find that the charged mode for  $\mu(B \to K^\ast \nu \bar{\nu})$ is the almost same equation as Eq.~\eqref{eq:BtoKst}.
}
Up to the CKM parameters, we find that these values are consistent with Refs.~\cite{Allwicher:2023xba,Gao:2024vql,Crivellin:2025qsq,Altmannshofer:2025eor,Das:2025zrn}.

The predicted signal strengths are theoretically correlated via the $b \to s \nu_i 
\bar{\nu}_j$ WCs in the effective Hamiltonian.

 %%%%%%%%%%%%%
\section{Exact Baryon-Meson Relation}\label{sec:SR}
%%%%%%%%%%%%%

From the expressions given in the previous section, 
we arrive at the following very compact formula among the three signal strengths 
(normalized to the SM predictions): 
\beq
\mu\left(\Lambda_b \to \Lambda \nu \bar\nu\right)
&= \left(0.26 \pm 0.05\right) \mu\left(B \to K \nu \bar{\nu}\right) \nonumber 
\\ & \quad + \left(0.74 \mp 0.05 \right)
 \mu\left(B \to K^\ast \nu \bar{\nu}\right)\,,
 \label{eq:SR}
\eeq
where all dependence on 18 WCs is identical on both sides.
The quoted uncertainties in the two coefficients arise from the form factors and are fully anti-correlated.
As discussed below, one can derive this formula a straightforward way. 
The algebraic origin of the relation becomes transparent once the signal strengths are rewritten in terms of the two independent WC 
combinations introduced below.

Due to the parity conservation in  QCD, $\left\langle\bar{K}\left(p_K\right)\right| \bar{s} \gamma_\mu \gamma_5 b\left|\bar{B}\left(p_B\right)\right\rangle =0$ holds in the hadronizations. 
This leads to \cite{Bause:2023mfe}
\beq
3 \mu \left(B \to K \nu \bar{\nu}\right) = \sum_{i}\sum_{j}| C_{+}^{ij}|^2\,,\label{eq:BK}
\eeq
with $
C_{\pm}^{ij} \equiv \delta^{ij}+C_L^{i j} \pm C_R^{i j}$.
On the other hand, 
although the vector and baryonic channels are more involved \cite{Hiller:2013cza}, their expressions can still be written in the form of 
\beq
\! 3 \mu\left(B \to K^\ast \nu \bar{\nu}\right)& = \sum_{i}\sum_{j} \left( a_+ | C_{+}^{ij}|^2 + a_- | C_{-}^{ij}|^2\right)\,,\\
3 \mu\left(\Lambda_b \to \Lambda \nu \bar{\nu}\right)& = \sum_{i}\sum_{j} \left( b_+ | C_{+}^{ij}|^2 + b_- | C_{-}^{ij}|^2\right)\,,
\label{eq:lambdab}
\eeq
with $a_+ = 0.20, a_- = 1- a_+ =  0.80, b_+ = 0.41,$ and $b_- = 1 - b_+ = 0.59$. 
The interference between $C_+$ and $C_-$ is absent after integrating over the unobserved neutrino kinematics.
These three equations immediately lead to the identity
\beq
\mu\left(\Lambda_b \to \Lambda \nu \bar{\nu}\right)
&= \frac{ b_+ - a_+}{a_- } 
\mu \left(B \to K \nu \bar{\nu}\right)
\nonumber \\
&\quad  + \frac{b_-}{a_- }\mu\left(B \to K^\ast \nu \bar{\nu}\right)\,.\label{eq:sm}
\eeq
Here, all dependence on the WCs has been eliminated.
This relation is algebraically exact within the effective Hamiltonian of Eq.~\eqref{eq:Heff},
while its coefficients depend on hadronic form factors and therefore carry uncertainties in their numerical evaluation.
Including the form factor uncertainties, details of their treatment are given in the Appendix \ref{sec:app}, one can derive the relation in Eq.~\eqref{eq:SR}.
In this way, this identity does not depend on whether neutrino flavor is conserved or violated.

Interestingly, 
we find that the central values of the coefficients for the pseudoscalar- and vector-meson modes in the 
relation~\eqref{eq:SR} are close to $1/4$ and $3/4$, respectively, which appear in the $b \rightarrow c$ semileptonic sum rule in the heavy-quark limit \cite{Endo:2025fke,Endo:2025lvy}. 
This numerical agreement should not, however, be interpreted as implying a common spin-counting origin. 
In the heavy-to-heavy $(b \rightarrow c)$ transitions, the pseudoscalar $(D)$ and vector-meson $(D^\ast)$ form a heavy-quark spin multiplet,
and the coefficients $1/4$ and $3/4$ can be understood directly 
from heavy-quark spin symmetry \cite{Isgur:1989vq,Isgur:1990yhj} and the corresponding spin counting.
By contrast, $K$ and $K^\ast$ do not form a heavy-quark spin multiplet.
At large recoil, 
the form factors
in the heavy-to-light $(b \rightarrow s)$ transitions are related within the corresponding channels, 
but are not unified into the independent soft form factors governing $B \rightarrow K$ and $B \rightarrow K^\ast$.
The present $b \rightarrow s$ relation instead follows from the explicit two-dimensional Wilson-coefficient dependence in Eqs.~\eqref{eq:BK}--\eqref{eq:lambdab}.

Mathematically, 
this follows from the fact that no three linearly independent vectors can exist in a two-dimensional vector space. 
This should be contrasted with generic $b \to c$ semileptonic sum rules,
where residual terms remain,
but the resulting relation still provides a good approximation.
In the $b \to s \nu_i\bar{\nu}_j$ observables,
this space is spanned by two summations of the WCs: $\sum_{i,j}|C_+^{ij}|^2$ and $\sum_{i,j}|C_-^{ij}|^2$.
Hence, any observable whose dependence on the WCs lies in this two-dimensional space obeys a similar exact 
relation, irrespective of neutrino flavor violation.
Thus, in a similar analysis, we derive another exact consistency relation
from Eq.~\eqref{eq:Lambdab} (see also \cite{Buras:2014fpa}), 
\beq
1 &= \left( 0.20 \pm 0.02 \right) \frac{\mu\left(B \to K \nu\bar\nu\right)}{\mu\left(B \to K^\ast \nu\bar\nu\right)} \nonumber \\
& \qquad + \left( 0.80  \mp 0.02\right) \mu\left(F_L(K^\ast)\right) \,,
\label{eq:SRFL}
\eeq
with anti-correlation between the uncertainties of the two terms. 
Again, all dependence on 18 WCs is canceled in the right-hand side.

Along similar lines, one can also construct a known relation between the inclusive and exclusive decays \cite{Altmannshofer:2009ma,Buras:2014fpa}
\beq
\mu\left(B \to X_s \nu\bar\nu\right)
&= \left(0.34 \pm 0.01\right) \mu\left(B \to K \nu \bar{\nu}\right) \nonumber 
\\ & \quad + \left(0.66 \mp 0.01 \right)
 \mu\left(B \to K^\ast \nu \bar{\nu}\right)\,,
 \label{eq:SRinc}
\eeq
where $\mu\left(B \to X_s \nu\bar\nu\right) = \mathcal{B}(B \to X_s \nu\bar\nu)/ \mathcal{B}(B \to X_s \nu\bar\nu)_{\rm SM}$ 
and the uncertainties are 
fully anti-correlated.
Interestingly again, 
we found that this expression is broadly consistent with the baryon-meson relation~\eqref{eq:SR}:
\beq
\mu(\Lambda_b \to \Lambda \nu \bar\nu) \approx 
\mu\left(B \to X_s \nu \bar{\nu}\right)\,.
\eeq
This intriguing agreement was already pointed out in the study of the $b \to c$ semileptonic sum rule  \cite{Blanke:2019qrx,Endo:2025set,Iguro:2026xgi}.

%%%%%%%%%%%%%%%%%%%
\begin{figure}[t]
\includegraphics[width=0.48\textwidth]{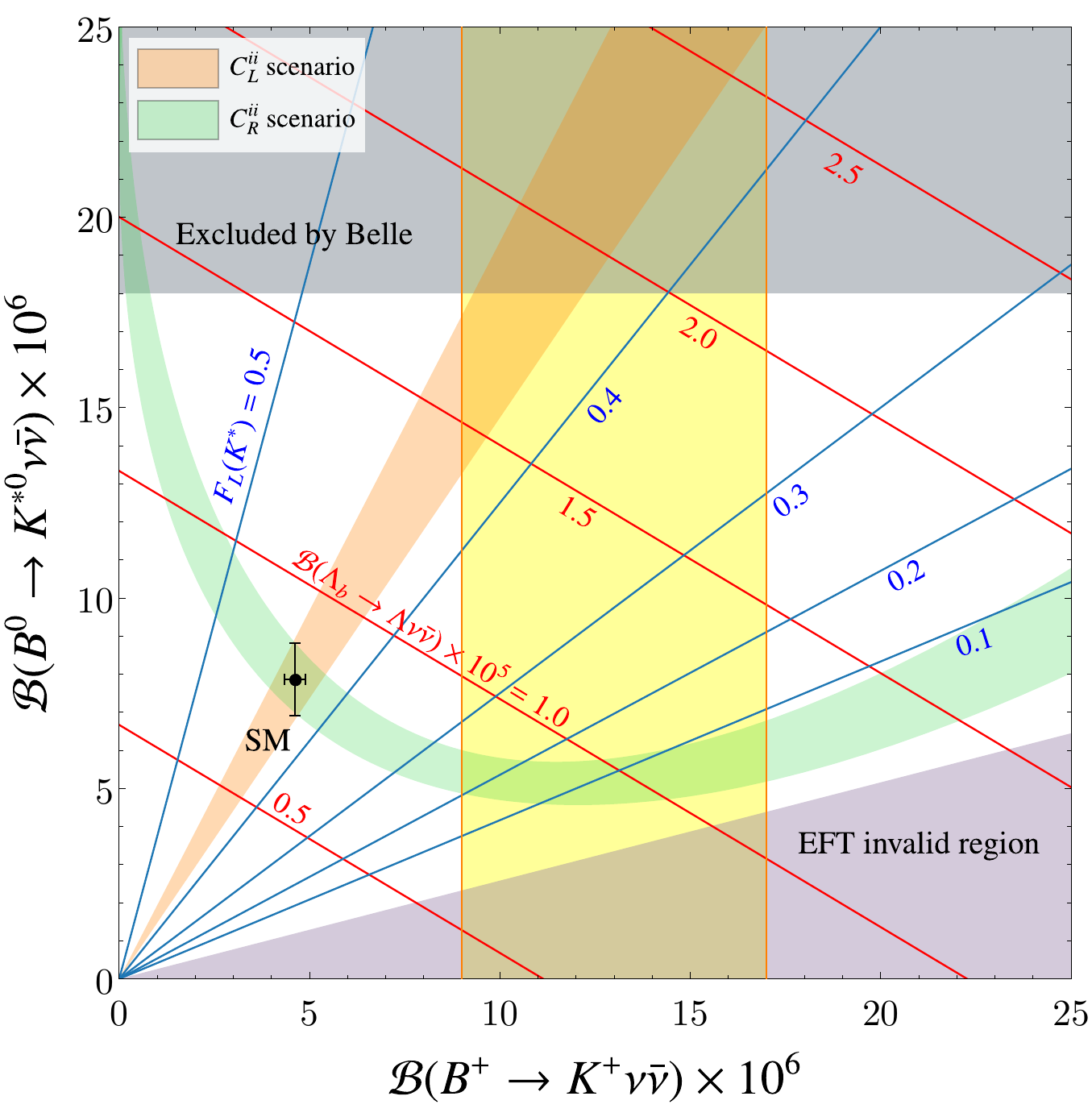}
\vspace{-0.3cm}
\caption{The contours of $\mathcal{B}(\Lambda_b\to \Lambda\nu\bar{\nu})\times10^5$ and 
$F_L(K^\ast)$
are shown as red and blue lines, respectively, 
on $\mathcal{B}(B^+\to K^+\nu\bar{\nu})$ and $\mathcal{B}(B^0\to K^{\ast 0}\nu\bar{\nu})$ plane. 
The orange and green regions represent the pure $C_L^{ii}$ and $C_R^{ii}$ scenarios, respectively, 
including the $1\sigma$ form factor uncertainties.
}
%\vspace{-0.3cm}
\label{fig:BR}
\end{figure}
%%%%%%%%%%%%%%%%%%%%%

In this manner, the observables of $\mathcal{B} (\Lambda_b \to \Lambda \nu \bar{\nu})$ and $F_L(K^\ast)$ 
can be indirectly determined from the measurements of $\mathcal{B}(B \to K^{(*)}\nu \bar{\nu})$ in a model-independent way. 
This relation connects among the baryonic and mesonic channels in a theoretically clean way and can provide a complementary probe of possible new physics effect. 
In particular, if future measurements establish a violation of these relations, 
this would indicate the presence of new physics beyond the framework with only left-handed neutrinos.

In Fig.~\ref{fig:BR}, 
following the argument in Ref.~\cite{Allwicher:2023xba}, 
we show the correlation among $b \to s \nu\bar\nu$ observables. 
For $\mathcal{B}(\Lambda_b\to \Lambda\nu\bar{\nu})$ and $F_L(K^\ast)$, we use the relations \eqref{eq:SR} and \eqref{eq:SRFL} without taking the theoretical uncertainties. 
The orange and green  colored regions indicate predictions of the pure $C_L^{ii}~(C_L^{11}=C_L^{22}=C_L^{33}\neq 0$ and the others are zero) and pure $C_R^{ii}$ scenarios, respectively.
The yellow-shaded region is the current data of $\mathcal{B}(B^+ \to K^+\nu\bar\nu)$ in Eq.~\eqref{eq:BelleIIresult}.
The gray-shaded region is excluded at 90\% CL by the Belle $B^0 \to K^{\ast 0} \nu\bar\nu$ search in Eq.~\eqref{eq:Kstbound}.

Using $|C_-^{ij}|^2 \geq 0$, one can readily derive the EFT bound
\cite{Bause:2023mfe,Allwicher:2023xba}:
\beq
& \mathcal{B}\left(B^0 \to K^{\ast 0} \nu \bar{\nu}\right) \nonumber \\
& \quad \geq a_+ \frac{\mathcal{B}\left(B^0 \to K^{\ast 0} \nu \bar{\nu}\right)_{\rm SM}}{\mathcal{B}\left(B^+ \to K^+ \nu \bar{\nu}\right)_{\rm SM}}
\mathcal{B}\!\left(B^+ \to K^+ \nu \bar{\nu}\right)\nonumber\\
&\quad  = (0.35 \pm 0.05) \,\mathcal{B}\!\left(B^+ \to K^+ \nu \bar{\nu}\right)\,.
\eeq

For small $\mathcal{B}(B^0\to K^{\ast 0}\nu\bar{\nu})$, 
we show the theoretically excluded parameter region by the EFT bound (at the $2\sigma$ level) as the dark-purple-shaded region.

%%%%%%%%%%%%%%%%%%%
\begin{figure}[t]
\includegraphics[width=0.48\textwidth]{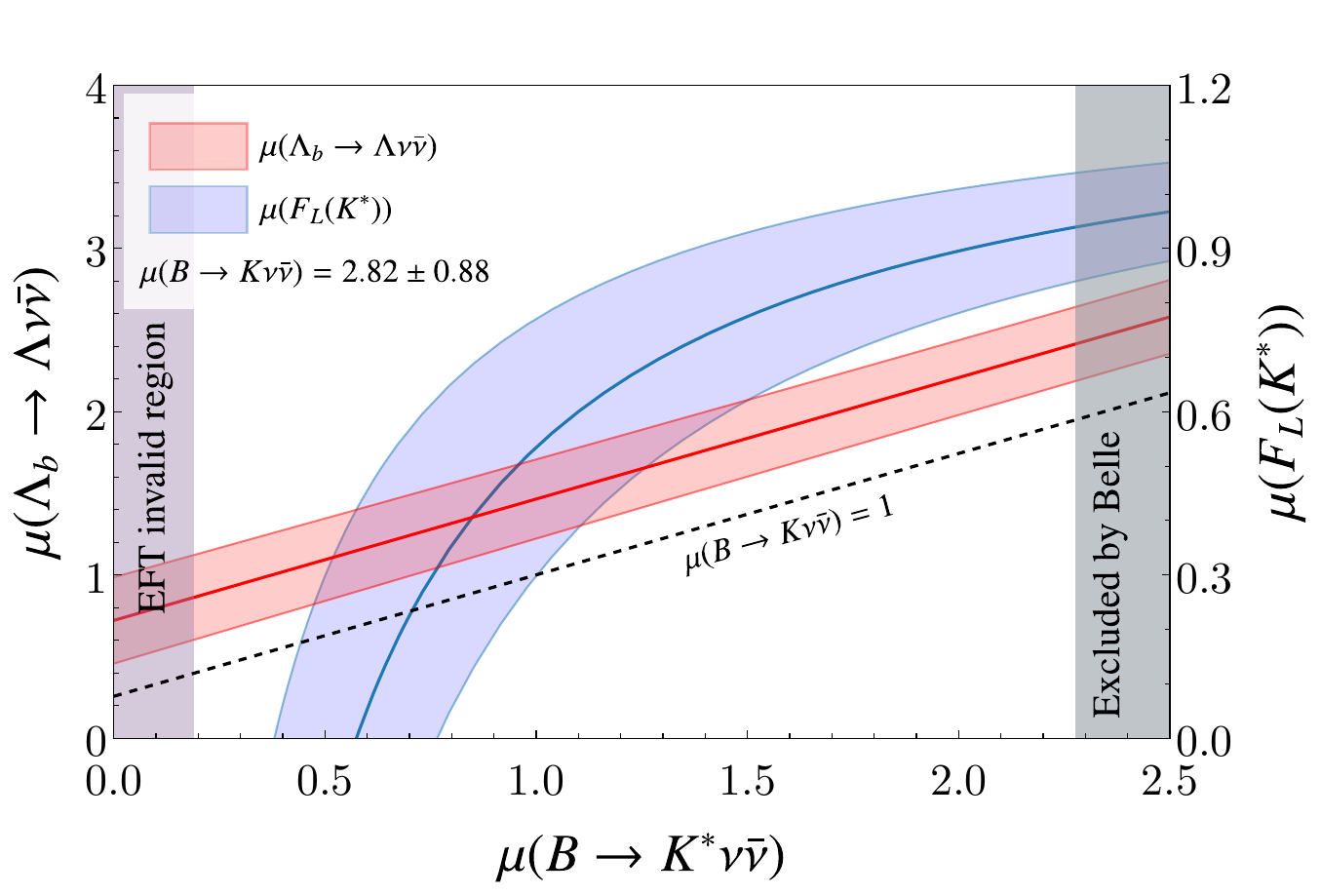}
\vspace{-0.3cm}
\caption{
Prediction of the baryon-meson relation using the Belle II result,
$\mu(B \to K \nu \bar{\nu})_{\rm exp}=2.82 \pm 0.88$. 
The red band represents $\mu(\Lambda_b\to\Lambda\nu\bar{\nu})$ labeled on the left axis, whereas the blue region represents $\mu(F_L(K^*))$ labeled on the right axis,
with 
$1\sigma$ uncertainties obtained from the form factors and $\mu(B \to K \nu \bar{\nu})_{\rm exp}$.
The black dashed line is a prediction for $\mu(B \to K \nu \bar{\nu})=1$.}
\label{fig:mu}
\end{figure}
%%%%%%%%%%%%%%%%%%%%%
To gain further insight into this correlation, 
Fig.~\ref{fig:mu} shows the corresponding  plot including  all uncertainties, 
where the value of $\mu(B \to K \nu \bar\nu)$ is fixed
to the latest data in Eq.~\eqref{eq:BelleIIresult}.
Predicted $\mu(\Lambda_b \to \Lambda \nu\bar\nu)$ and $\mu(F_L(K^\ast))$ are shown by the red and blue regions, respectively.
We find that these uncertainties are currently dominated by $\mu(B \to K \nu \bar\nu)_{\rm exp}$, while the form factor uncertainties also give a sizable contribution.
Again, the gray-shaded region
and
dark-purple-shaded region are excluded by the Belle and the EFT bounds, respectively.  

It is clearly shown that once $B \to K^\ast \nu \bar\nu$ is measured, the branching fraction of $\Lambda_b \to \Lambda \nu\bar\nu$ and the polarization fraction $F_L(K^\ast)$ can be predicted rather concretely and tested in future measurements.
In particular, the current data already sets the upper bound
$\mu(\Lambda_b \to \Lambda \nu \bar\nu) \leq 2.6 \,(2.9)$ at the $1\sigma\,(2\sigma)$ level.

%%%%%%%%%%%%%%%%%%%%%%%%%%%%%%%%%%%%%
\section{Discussion and Conclusions}{\label{sec:conc}}
%%%%%%%%%%%%%%
In this paper,
we have established the robust relation among the branching fractions of $\Lambda_b \to \Lambda \nu \bar\nu$ and $B \to K^{(\ast)} \nu\bar\nu$, assuming decoupled right-handed neutrinos.
We find the fact that this relation remains exact even in the presence of 18 independent WCs can be understood from 
the explicit structures in Eqs.~\eqref{eq:BK}--\eqref{eq:sm}.
As a result, 
once the decay rate of $B \to K^{\ast} \nu \bar\nu$ is measured at Belle II,
that of $\Lambda_b \to \Lambda \nu\bar\nu$ 
can be inferred immediately
in a model-independent way within new-physics scenarios 
involving only active neutrinos. 

The present relation is algebraically exact within the effective Hamiltonian of Eq.~\eqref{eq:Heff}, 
while its coefficients depend on hadronic form factors and carry uncertainties in their numerical evaluation.
Consequently, any significant experimental deviation from this relation would provide evidence for physics beyond this framework, 
such as additional light dark particles \cite{He:2023bnk,Fridell:2023ssf,Gabrielli:2024wys,Hou:2024vyw,He:2024iju,Bolton:2024egx,Bolton:2025fsq},  light right-handed neutrinos \cite{Bause:2023mfe,Das:2025zrn}, or lepton-number violation \cite{Felkl:2021uxi,Buras:2024ewl,Kim:2024tsm,Endo:2026qof}.
This result thus establishes $b \to s \nu \bar\nu$ observables as a powerful tool for discriminating among new physics scenarios in future experiments.
How significantly these effects violate this relation will be investigated in future work.

We have also pointed out that the coefficients of this baryon-meson relation are numerically close to those of the $b\to c$ semileptonic sum rule.
Although this numerical agreement is suggestive, 
a direct heavy-quark-symmetry derivation is not available for the heavy-to-light $b\to s$ case. 
In the present work, this relation  is derived instead from the explicit structure of the rates.
Related heavy-to-light
semileptonic sum rules have recently been discussed in charm decays \cite{Endo:2026blh}, where residual terms remain while the sum-rule relation can still
be phenomenologically useful.

It is worth emphasizing once again that, even in the absence of data for the baryon mode, 
the meson modes alone are sufficient to provide a meaningful validation in a model-independent way.
According to Ref.~\cite{Belle-II:2018jsg}, 
it is expected that Belle II can observe $B \to K^{\ast 0}\nu\bar{\nu}$ at early stage. 
Both sensitivities on the branching fractions of $B^+ \to K^+ \nu \bar{\nu}$ and $B^{0,+}  \to K^{\ast 0,+} \nu \bar{\nu}$ are 10\% level, and the one on $F_L(K^\ast)$ is $0.08$ with 50\,ab${}^{-1}$.
These sensitivities should be sufficient to enable an experimental verification of the 
exact consistency relation
for the polarization fraction in \eqref{eq:SRFL}, and hopefully also of the inclusive relation in \eqref{eq:SRinc} \cite{Belle-II:2025bho}.

Finally, the corresponding $b \to d \nu \bar{\nu}$ modes offer a natural extension of the $b \to s \nu \bar{\nu}$ analysis.
Although they are additionally CKM-suppressed in the SM,
Belle measurements of the mesonic channels have already established upper limits \cite{Belle:2017oht, Belle:2013tnz}. 
They therefore remain well-motivated targets for future searches, particularly in light of the significantly improved sensitivity expected at next-generation experiments.

\vspace{5pt}
%%%%%%%%%%%%%%%%%%%%%%%%%%%%%%%%%%%%%%%%%%%%%%%%%
{\it Acknowledgments} --- {\small 
We would like to thank Akimasa Ishikawa, Shohei Nishida, and Xunwu Zuo 
for helpful discussions from the experimental perspective.
We also thank Diganta Das, Anjan Giri, Hector Gisbert, 
Karthik Jain, Dargi Shameer, and  Takeru Uchiyama for useful exchanges. 
We thank Syuhei Iguro and Koki Tomita for helpful comments and valuable suggestions for improving the uncertainty analysis.
We would like to thank the organizers and participants of Workshop on High Energy Physics Phenomenology 2025 (WHEPP-2025), 
where stimulating discussions led to the initial idea for this work.
T.K.\ and K.S.\ also gratefully acknowledge Theory-Experiment Assembly workshop,
where this work was discussed.
The work of T.K.\ is supported by the JSPS Grant-in-Aid for Scientific Research Grant No.\,24K22872 and 25K07276. 
}
%%%%%%%%%%%%%%%%%%%%%%%%%%%%%%%%%%%%%%%%%%%%%%%%%
\appendix
%%%%%%%%%%%%%
\section{Uncertainty for the branching ratio of the $B\to M$ transition}
\label{sec:app}
%%%%%%%%%%%%%

In general, 
the form factors of the $B\to M~(M = K, K^\ast)$ transitions (and also $\Lambda_b \to \Lambda$ transition) are parameterized by the 
BSZ
$z$-series expansion
\cite{Bharucha:2015bzk} (BSZ-type expansion for the $\Lambda_b \to \Lambda$ decays \cite{Detmold:2016pkz}):
\beq
\mathcal{F}_{B\to M}^i(q^2)=\frac{1}{1-\frac{q^2}{m_{i, \text{pole}}^2}}\sum_{n=0}^{N-1} \alpha_n^i\left[z(q^2)-z(0)\right]^n \,,
\eeq
with
\beq
z(q^2) =\frac{\sqrt{t_+-q^2}-\sqrt{t_+-t_0}}{\sqrt{t_+-q^2}+\sqrt{t_+-t_0}}\,,
\label{eq:z0}
\eeq
where the label $i$ denotes the type of form factors,
$m_{i, \text{pole}}$ is the mass of the lowest-lying resonance and
$t_\pm=(M_B\pm M_M)^2$, $t_0=t_+ - \sqrt{t_+ (t_+-t_-)}$.
(For the $\Lambda_b \to \Lambda$ decays, $t_0 = (m_{\Lambda_b} - m_{\Lambda})^2 = q^2_{\rm max}$, and $z(0)$ in Eq.~\eqref{eq:z0} should be $z(q^2_{\rm max})$.)
In our analysis,
we truncate the $z$-series expansion at $N=3$, since $|z(q^2)|_\text{max}\leq 0.1$. 
The $z$-series coefficients $\alpha_n^i$, together with their uncertainties and covariances, 
are determined from lattice QCD, dispersive bounds, and light-cone sum rules.

Below, we briefly summarize how the uncertainties in $\alpha_n^i$ propagate to the branching ratio in our analysis.
Since the branching ratios $\mathcal{B}$ are a quadratic form of the form factors, 
they can formally be written in the following form:
\begin{equation}
\mathcal{B}
=\sum_{i,j} \alpha_i C_{ij} \alpha_j\,,
\end{equation}
where $\alpha_i$ denotes the coefficient vector with components $\alpha_n^i$ (we relabeled the index $i$).
The matrix $C_{ij}$ consists of entries given by functions of the WCs after performing the momentum ($q^2$) integration, with 
 $C_{ij}=C_{ji}$. 
 The covariance of the vector $\alpha_i$ is given as the correlation matrix $\rho_{ij}$, defined by
\beq
\text{Cov}(\alpha_i,\alpha_j)=
\sigma_i \rho_{ij} \sigma_j\,, 
\eeq
where $\sigma_i$ denotes the uncertainty of $\alpha_i$ with $\langle\sigma_i \rangle=0$. Here,
any matrix element $\rho_{ij}$ satisfies  $-1\leq\rho_{ij}\leq 1$ and $\rho_{ii}=1$.

We use the correlation matrix $\rho_{ij}$ for the decays of $B \to K \nu \bar{\nu}$ \cite{Gubernari:2023puw}, 
$B \to K^* \nu \bar{\nu}$~\cite{Gao:2024vql} and 
$\Lambda_b \to \Lambda \nu \bar{\nu}$~\cite{Detmold:2016pkz}, 
together with the uncertainties of the coefficients $\alpha_i$, 
to calculate the uncertainties of the branching ratios.

To propagate the form factor uncertainties, 
we consider a shift in the parameters
$\alpha_i \rightarrow \alpha_i + \sigma_i$. 
Expanding the branching ratio $\mathcal{B}$ to first order then gives
\begin{equation}
\mathcal{B} \rightarrow \mathcal{B} 
+ \Delta\mathcal{B} \,,
\qquad 
\Delta \mathcal{B} \simeq \sum_i \frac{\partial \mathcal{B}}{\partial \alpha_i}\,\sigma_i \,.
\end{equation}
The variance of $\mathcal{B}$ can be obtained by
\begin{align}
\text{Var}(\mathcal{B}) 
&= \langle (\Delta \mathcal{B})^2 \rangle  \nonumber \\
&\simeq 
\left\langle 
\left( \sum_j \frac{\partial \mathcal{B}}{\partial \alpha_j} \sigma_j\right) 
\left( \sum_k \frac{\partial \mathcal{B}}{\partial \alpha_k} \sigma_k\right) 
\right\rangle \nonumber \\
&=  \sum_{j,k} \frac{\partial \mathcal{B}}{\partial \alpha_j} 
\langle \sigma_j \sigma_k \rangle 
\frac{\partial \mathcal{B}}{\partial \alpha_k} \nonumber \\
&=: \sum_{j,k} \frac{\partial \mathcal{B}}{\partial \alpha_j} 
\text{Cov}(\alpha_j,\alpha_k)
\frac{\partial \mathcal{B}}{\partial \alpha_k} \nonumber \\
&= \sum_{j,k} 
\frac{\partial \mathcal{B}}{\partial \alpha_j} 
\sigma_j \rho_{jk} \sigma_k 
\frac{\partial \mathcal{B}}{\partial \alpha_k}\,,
\end{align}
which is consistent with Ref.~\cite{Bharucha:2015bzk}.
Therefore, we obtain
\begin{equation}
\begin{aligned}
\text{Var}(\mathcal{B})
&\simeq  \sum_{i,j,k,l}
\alpha_i (C_{ij}+C_{ji}) 
\sigma_j \rho_{jk}\sigma_k 
(C_{kl}+C_{lk}) \alpha_l\\
&=4 \sum_{i,j,k,l}
\alpha_i C_{ij} \sigma_j \rho_{jk}\sigma_k C_{kl} \alpha_l \,,
\end{aligned}
\end{equation}
here we used $C_{ij}=C_{ji}$. 
Eventually, the branching ratios including the form factor uncertainty from from factor are represented by
\begin{equation}
\mathcal{B}\pm\sqrt{\text{Var}(\mathcal{B})}\,.
\end{equation}
%%%%%%%%%%%%%%%%%%%%%%%%%%%%%%%%%%%%%%%%%%%%%%%%%
\bibliographystyle{utphys28mod}
\bibliography{ref}
%%%%%%%%%%%%%%%%
\end{document}